\documentclass[aps,prl,twocolumn,floats]{revtex4}
\usepackage{color,graphicx,pstricks}
\begin{document}

\title {The Distinct Effects of Homogeneous Weak Disorder and
Dilute Strong Scatterers\\
on Phase Competetion in the Manganites}

\author{Kalpataru Pradhan, Anamitra Mukherjee and Pinaki Majumdar}

\affiliation{Harish-Chandra  Research Institute,
 Chhatnag Road, Jhusi, Allahabad 211 019, India}

\date{11 Aug, 2007}

\begin{abstract}
We study the two orbital double-exchange model in two dimensions in the
presence of antiferromagnetic (AF) superexchange, strong Jahn-Teller
coupling, and substitutional disorder. At hole doping $x=0.5$  we explore
the `bicritical' regime where the energy of a ferromagnetic metal and a
charge and orbital ordered (CO-OO) CE state are closely balanced, and
compare the impact of weak homogeneous disorder to that of a low density
of strong scatterers. Even moderate homogeneous disorder suppresses the
CE-CO-OO phase and leads to a glass with nanoscale correlations. Dilute
strong scatterers of comparable strength, however, convert the CE-CO-OO
phase to a {\it phase separated} state with ferromagnetic and AF-CO-OO
clusters.  We provide the first spatial description of these phenomena
and compare our results in detail to experiments on the half-doped
manganites.

\end{abstract}

\maketitle

The manganese oxides of the form  A$_{1-x}$A'$_x$MnO$_3$
involve a remarkable interplay of charge, spin, lattice,
and orbital degrees of freedom \cite{mang-book}. 
This cross coupling is
most striking in the half doped $(x=0.5)$ manganites
many of which have a charge and orbital ordered insulating 
(CO-OO-I) ground state with `CE' magnetic order 
- a zigzag pattern of 
ferromagnetic chains with antiferromagnetic 
(AF) coupling between them.  The CE-CO-OO-I phase shows up in 
manganites with low mean cation radius $(r_A)$ while 
systems with large $r_A$ are ferromagnetic metals (FM-M).
The  variation of $r_A$ leads to a `bicritical' phase
diagram \cite{ce-bicr1}
with a first order boundary between the 
FM-M and the CE-CO-OO-I phases. 

Disorder has a remarkable effect on the bicriticality.
Even moderate `alloy' disorder,
due to random location of A and A' ions 
at the rare earth site, 
converts the CO-OO-CE phase to 
a short range correlated glass,
but has only limited impact on the 
ferromagnet \cite{ce-bicr1,ce-bicr2,ph-comp-rev}.
The asymmetric suppression of spatial order by
cation disorder and 
the emergence of a charge-orbital-spin glass at low $r_A$
are one set of intriguing 
issues in these materials. Unusually,
while alloy type randomness on the A site
leads to a {\it homogeneous glassy phase}, 
the substitution of a few percent of  Mn (the `B site')
by Cr  \cite{cr-old1,cr-old2} leads
to {\it phase separation} of the system 
\cite{cr-tok1,cr-tok2,cr-ps1,cr-ps2} 
into FM-M and AF-CO-OO-I domains.
The difference between A and B site disorder
holds the key to the much discussed phase coexistence
and spatial inhomogeneity in the manganites.

In this paper  we provide the first results on the 
relative effects of A and B type 
substitutional disorder on phase competetion 
in a manganite model. We study 
weak `alloy' disorder and dilute strongly 
repulsive  scatterers. Our main results are:
(i)~Alloy disorder indeed 
leads to asymmetric suppression of long range order; 
moderate disorder converts long range 
CE-CO-OO to an {\it insulating glass}
with nanoscale inhomogeneities, while 
FM order is only weakened.
(ii)~A low density, $\gtrsim 4\%$, of strong scatterers 
in the CE phase 
leads to cluster coexistence of AF-CO-OO and 
FM regions  and the ground state is a {\it poor metal}.
(iii)~The impact of strong scatterers depends crucially
on whether they are attractive or repulsive,  it 
correlates
with the asymmetry of the `clean' system about
$x=0.5$, and uncovers a new route for phase control.

We consider a two band model for $e_g$
electrons, Hunds coupled to $t_{2g}$ derived core spins, in
a two dimensional square lattice. The electrons are also
coupled to Jahn-Teller phonons, while the core spins have
an AF superexchange coupling between them.
These ingredients are all necessary to obtain a 
CE-CO-OO phase.
We include the effect of disorder through an 
on site potential.
\begin{eqnarray}
H &=& \sum_{\langle ij \rangle \sigma}^{\alpha \beta}
t_{\alpha \beta}^{ij}
 c^{\dagger}_{i \alpha \sigma} c^{~}_{j \beta \sigma} 
+ \sum_i (\epsilon_i -\mu)n_i 
~ - J_H\sum_i {\bf S}_i.{\mbox {\boldmath $\sigma$}}_i \cr
&&
~~+ J_{AF}\sum_{\langle ij \rangle}
{\bf S}_i.{\bf S}_j
 - \lambda \sum_i {\bf Q}_i.{\mbox {\boldmath $\tau$}}_i
+ {K \over 2} \sum_i {\bf Q}_i^2. 
\end{eqnarray}
\noindent
Here, $c$ and $c^{\dagger}$ are annihilation and creation operators for
$e_g$ electrons and
$\alpha$, $\beta $ are the two Mn-$e_g$ orbitals
$d_{x^2-y^2}$ and $d_{3z^2-r^2}$, 
labelled $(a)$ and $(b)$ in what follows.~$t_{\alpha \beta}^{ij}$~are 
hopping amplitudes between
nearest-neighbor sites with the
symmetry dictated form:~$t_{a a}^x= t_{a a}^y \equiv t$,~$t_{b b}^x= t_{b b}^y 
\equiv t/3 $,~$t_{a b}^x= t_{b a}^x \equiv -t/\sqrt{3} $,~$t_{a b}^y= t_{b a}^y 
\equiv t/\sqrt{3} $,~where
$x$ and $y$ are spatial directions 
We consider effectively a lattice of Mn ions and treat 
the alloy disorder due to 
cationic substitution 
as a random
potential $\epsilon_i$ at the Mn site 
picked from the distribution
$P_A(\epsilon_i)
= {1 \over 2}(\delta(\epsilon_i - \Delta) + 
\delta(\epsilon_i + \Delta))$.
The  Cr doping case is modelled via
$P_B(\epsilon_i)
= \eta \delta(\epsilon_i - V) + (1-\eta) \delta(\epsilon_i)$,
where $\eta$ is the percent substitution and $V$ the 
effective potential at the impurity site.
The $e_g$ electron spin is 
${\sigma}^{\mu}_i=
\sum_{\sigma \sigma'}^{\alpha} c^{\dagger}_{i\alpha \sigma}
\Gamma^{\mu}_{\sigma \sigma'}
c_{i \alpha \sigma'}$, where the
 $\Gamma$'s are Pauli matrices.
It is coupled to the
$t_{2g}$ spin ${\bf S}_i$ via the Hund's coupling
$J_H$, and we assume  $J_H/t \gg 1$.
$\lambda$ is the coupling between the JT distortion
${\bf Q}_i = (Q_{ix}, Q_{iz})$ and
the orbital pseudospin
${\tau}^{\mu}_i = \sum^{\alpha \beta}_{\sigma}
c^{\dagger}_{i\alpha \sigma}
\Gamma^{\mu}_{\alpha \beta} c_{i\beta \sigma}$, and
$K$ is the lattice 
stiffness.
We set $t=1$,   $K=1$,
and treat the ${\bf Q}_i$ and ${\bf S}_i$ as classical
variables \cite{class-approx}. The chemical potential $\mu$
is adjusted so that the electron density 
remains  $n=1/2$ which is
also $x= 1-n =1/2$.
For  A type disorder
the mean value is 
${\bar \epsilon_i }= 0$ and the variance is
$\Delta_A^2 = \langle (\epsilon_i 
- {\bar \epsilon_i })^2  \rangle = \Delta^2$, 
while for B type disorder 
${\bar \epsilon_i }=
 \eta V$ and $\Delta_B^2 =
\langle (\epsilon_i
- {\bar \epsilon_i })^2  \rangle~=~V^2 \eta(1-\eta)$.

The clean CE ground state at $x=0.5$ has been studied earlier
\cite{brink-prl,dag-ce-prl,brey-prb,cepas,dong-jt-prb}
using mean field and Monte Carlo (MC) techniques and is
well understood.
The impact of disorder on 
the phase competetion 
appropriate to $x=0.5$ has been studied on small clusters
\cite{ce-dag-03,motome-prl,ce-dis-dag}
usually using simplified models either without orbital
variables \cite{motome-prl} or ignoring the
electron-phonon coupling \cite{ce-dis-dag}.
The difficulty of simulating the full model, eqn[1], on
a large system has prevented any conclusive 
study.
We use our
recently developed travelling cluster approximation (TCA)
based MC 
\cite{tca}
to solve the problem.
Compared to exact diagonalisation (ED) based
MC which can handle typical sizes $\sim 8 \times 8$ 
we study the full model on lattices upto 
$40 \times 40$. In all our studies we use a moving
cluster of size $\sim 8 \times 8$ \cite{tca} to anneal
the spin and phonon  variables.

Before discussing the effect of disorder we determine
the clean ground
state at $x=0.5$ for varying $J_{AF}$ and $\lambda$, 
Fig.1.(a).

\begin{figure}
\vspace{.2cm}
\centerline{
\includegraphics[width=8.2cm,height=4.2cm,clip=true]{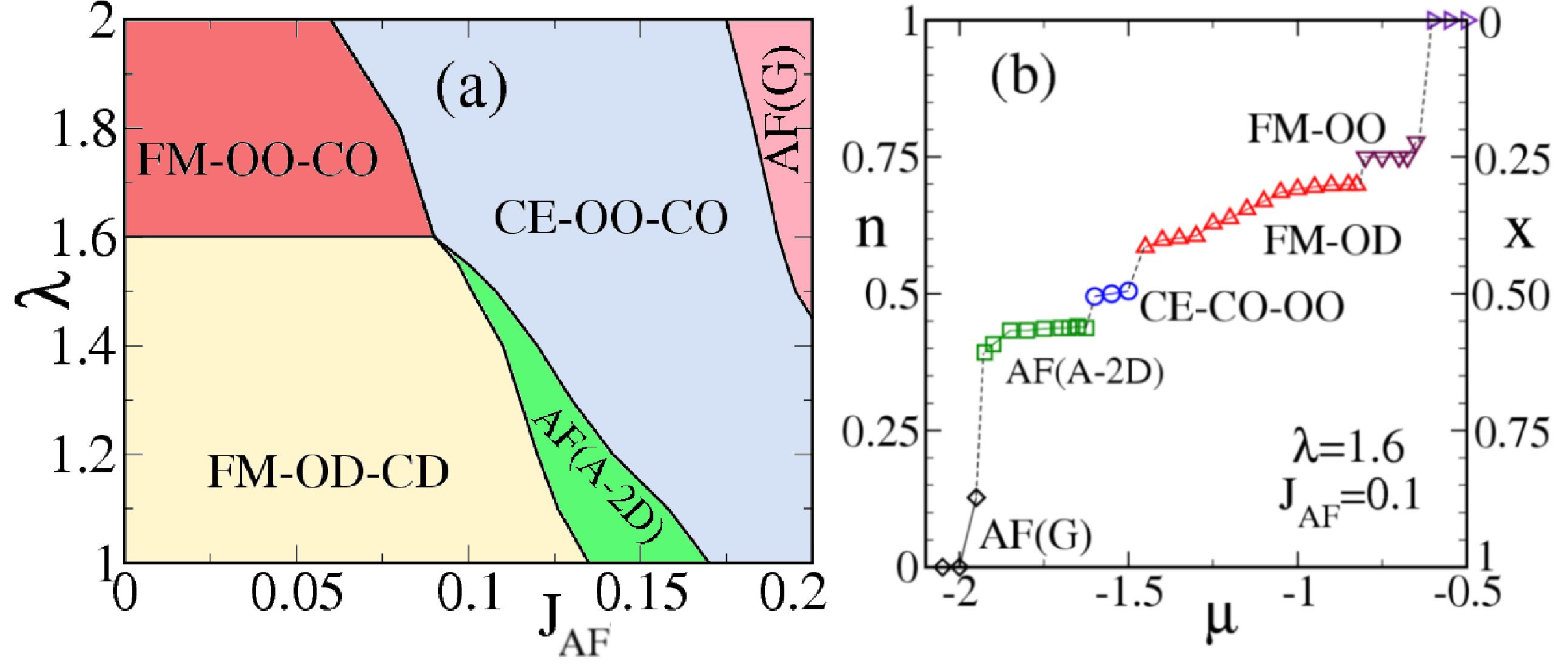}}
\vspace{.2cm}
\caption{Colour online: (a)~The ground state at $x=0.5$ for
varying $J_{AF}$ and $\lambda$, in
the absence of disorder. (b)~The doping $(n=1-x)$
dependence of the
ground state for varying chemical potential $\mu$
and typical electronic couplings, $\lambda =1.6$ and $J_{AF}=0.1$,
near the FM-OD-CD \& CE-CO-OO phase boundary.
The phases in the vicinity of $x=0.5$ are expected to show up
in a cluster pattern on introducing disorder at~$x=0.5$.}
\vspace{.2cm}
\end{figure}
\begin{figure}
\centerline{
\includegraphics[width=8.6cm, height=5.2cm, clip=true]{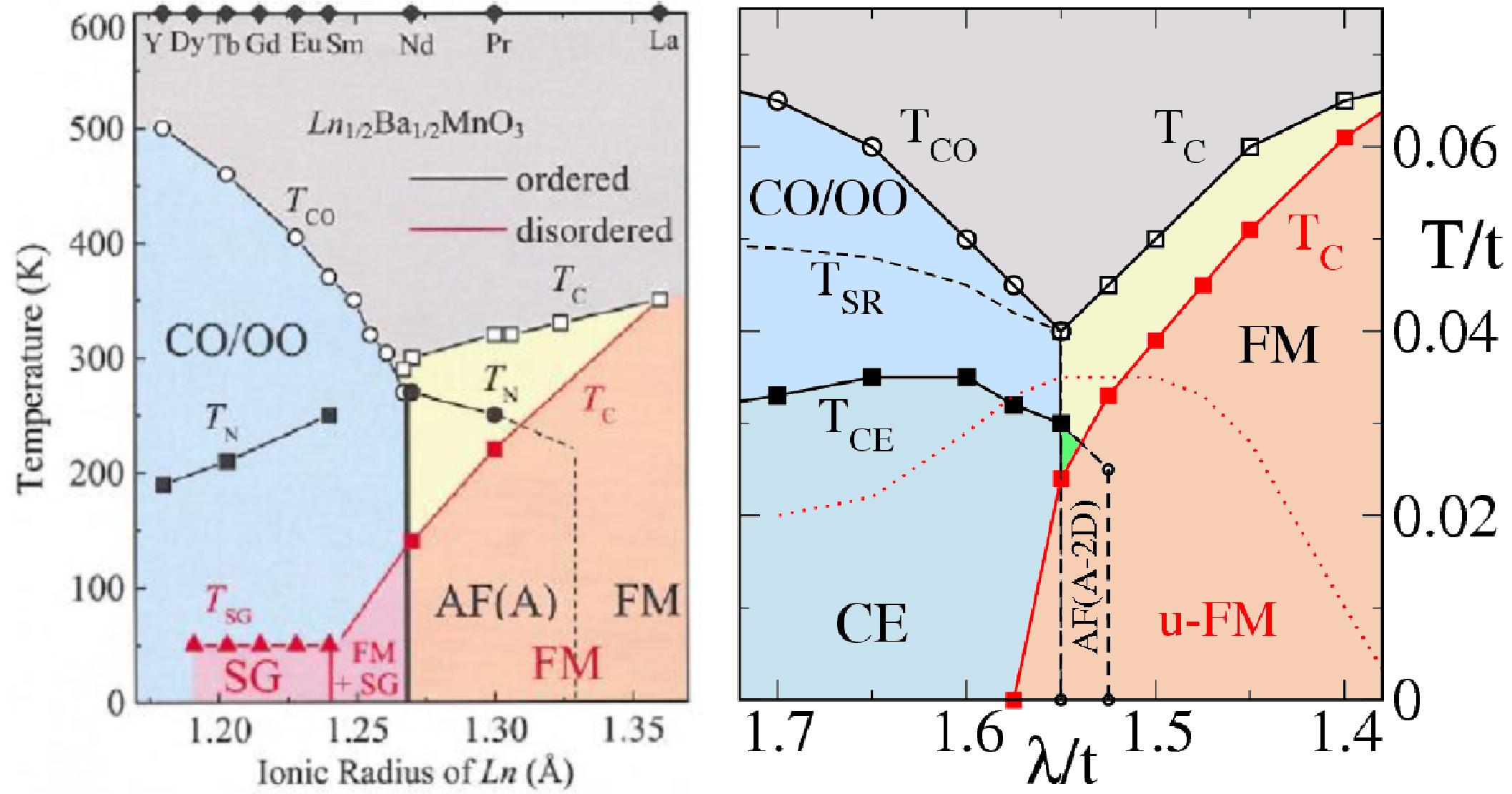}}
\vspace{.2cm}
\caption{Colour online: (a)~Experimental `bicritical'
 phase diagram in the
$x=0.5$ manganites obtained for ordered and
disordered (alloy) structures.
(b)~Our results: superposed
phase diagrams at $x=0.5$ for  $\Delta=0$ and
$\Delta =0.3$. The long range  CE-CO-OO
for $\lambda >  1.55$ at $\Delta=0$
is completely wiped
out at $\Delta=0.3$  while the FM-M phase at low
$\lambda$ becomes an unsaturated FM
with short range A-2D type correlations.
}
\vspace{.2cm}
\end{figure}
                                                             
At low $\lambda$ and low $J_{AF}$ double exchange is the
dominant interaction and kinetic energy optimisation leads
to a homogeneous ferromagnetic
state without any orbital or charge order
(FM-OD-CD). This phase has a finite density of states 
at the Fermi
level $\epsilon_F$ and is metallic. As $J_{AF}$ is
increased, keeping the JT coupling small, a magnetic state
emerges with   peaks in the structure factor
$S_{mag}({\bf q})$ at ${\bf q} = \{0, \pi\}$ or
$\{\pi, 0\}$ (we call this the A-2D phase), 
then an orbital ordered  but
uniform density
CE phase, with simultaneous peaks 
at ${\bf q} = \{0, \pi\},
\{\pi, 0\}$, and $\{\pi/2, \pi/2\}$. At even larger $J_{AF}$ 
the dominant correlations are `G type' with a peak 
at ${\bf q} = \{\pi, \pi\}$.
By contrast, 
increasing $\lambda$ at weak $J_{AF}$  keeps the
system ferromagnetic but leads to charge and orbital order
(FM-CO-OO) for $\lambda \gtrsim 1.6$.
Our interest is in a {\it charge ordered} CE phase. Such 
a state 
shows up when both  $\lambda$
and $J_{AF}$ are moderately large. 
The TCA based phase diagram is broadly 
consistent with previous variational results  
\cite{brink-prl,brey-prb,cepas,dong-jt-prb}
and with ED-MC on small systems \cite{dag-ce-prl}.

Since the effect of disorder might be to create cluster
coexistence \cite{dag-ps,nano-prl} of phases
{\it of different  densities}  that 
arise in the clean limit, Fig.1.(b) shows
 the phases and
phase separation windows that occur at a typical 
coupling, $J_{AF}=0.1$ and $\lambda=1.6$. For these
couplings the clean system is a CE-CO-OO phase
at $x=0.5$, a FM-M for $x \lesssim 0.4$,
 and an A-2D type AF for $x \gtrsim 0.55$.

To minimise the number of parameters, in
what follows we set $J_{AF}=0.1$.
This is in the right ballpark considering the AF transition
temperature at $x=1$, and allows close proximity of
the CE and FM-M phases.
We mimic the bandwidth
variation  arising from changing $r_A$
by varying $\lambda/t$  across the boundary between
CE-CO-OO and FM-OD-CD, and now explore the effects of
thermal fluctuation and disorder.

The key experiment \cite{ce-bicr1}
on the effect of A site disorder on 
bicriticality compared an  `ordered' 
structure, where the rare earth and alkaline earth ions
{\it sit on alternate layers}, with  
the `disordered' case where they are randomly distributed. 
The result is reproduced in the left panel in Fig.2.
While the ordered case has large transition temperatures for
the CO-OO, CE,  FM phases, {\it etc},
a random distribution of A and A' ions
destroy the CO-OO-CE phase and partially
suppresses the ferromagnetic~$T_c$.

\begin{figure}
\centerline{
\includegraphics[width=8cm, height=7.5cm, clip=true]{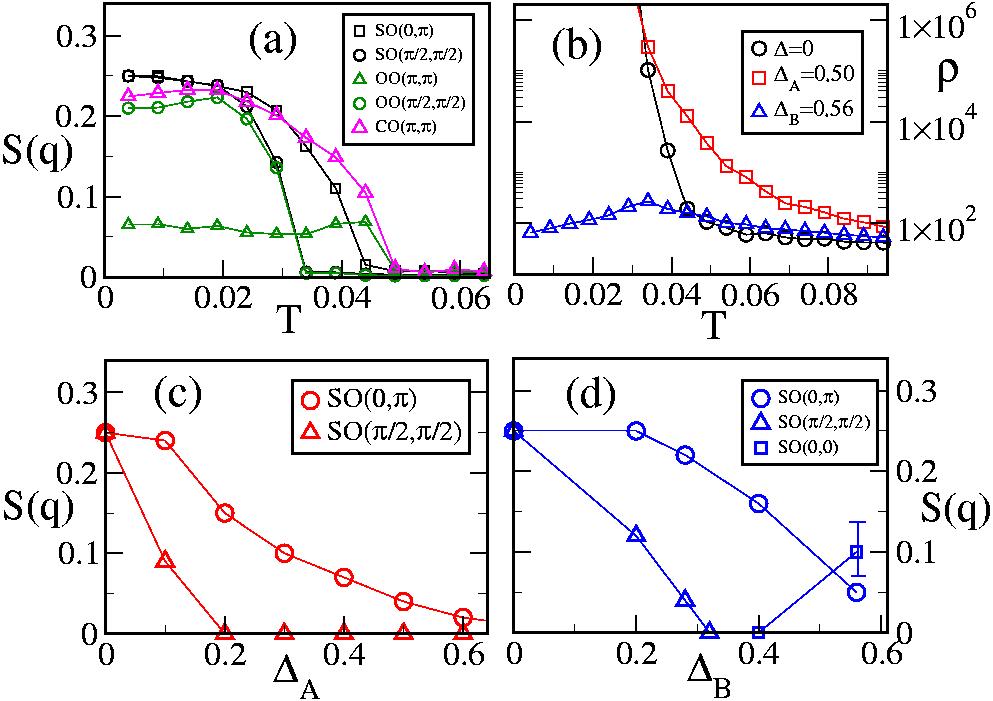}
}
\vspace{.2cm}
\caption{Colour online: Structure factors and resistivity
at $J_{AF}=0.1$ and  $\lambda/t =1.6$.
(a)~The $T$ dependence of the major peaks in the structure
factor for spin order (SO), orbital order (OO) and charge order
(CO) in the clean limit. Note the clear separation of scales
between $T_{CO}$, $T_{SR}$ and $T_{CE}$. (b)~The resistivity
$\rho(T)$ in the clean CE-CO-OO case and in the presence
of A type and B type disorder, with $\Delta_A \approx
\Delta_B \sim 0.5$. The $\Delta_B$ corresponds to $V=2$
and dilution $\eta=0.08$. (c)~Variation of the major
peaks in the magnetic structure factor with $\Delta_A$
at low temperature ($T=0.005$). (d)~Same as (c), now
with B type disorder, $V=2$ and varying $\eta$. Note the
emergence of the FM ${\bf q} =\{0, 0\}$ peak around
$\Delta_B =0.4$ ($\eta=0.04$). }
\vspace{.2cm}
\end{figure}
\begin{figure}
\centerline{
\includegraphics[width=8cm, height=7.9cm, clip=true]{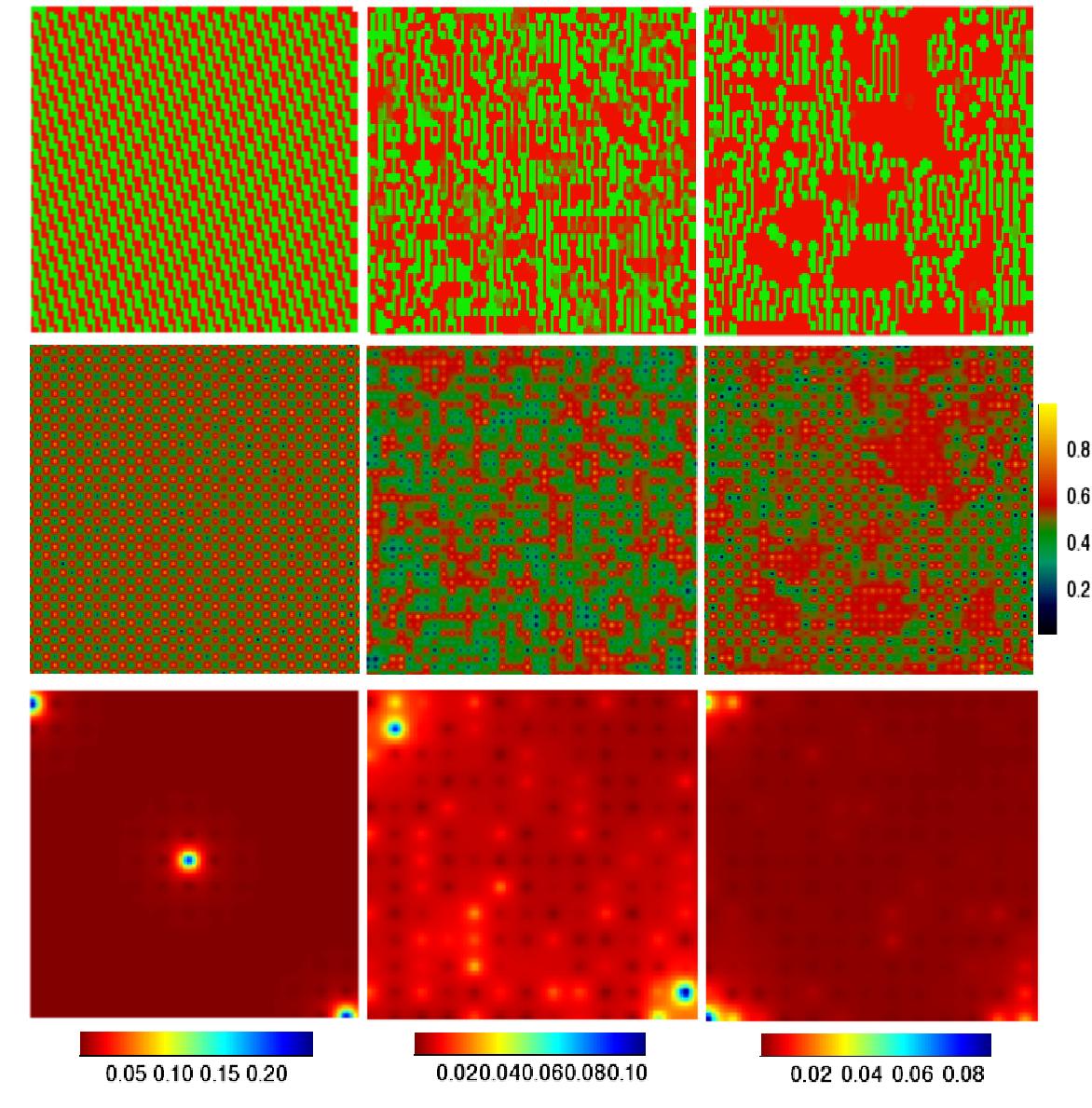}
}
\vspace{.2cm}
\caption{Colour online: MC snapshots and magnetic structure factor
at low temperature, $T=0.01$, size $40 \times 40$.
Left row: $\lambda=1.6$, non disordered, middle row, $\lambda=1.6$,
A type disorder with $\Delta_{eff}=0.5$, right row, $\lambda=1.6$,
B type disorder with $V=2$, $\eta= 8\%$, $\Delta_{eff} =0.56$.
Top panel shows the nearest neighbour magnetic correlation
${\bf S}_i.{\bf S}_{i + \delta}$, where $\delta = x$ or $y$.
Middle panel shows the charge density $\langle n_i
\rangle$ for the configuration
above. Bottom panels shows the MC averaged~$S_{mag}({\bf q})$.
In each panel ${\bf q} = \{0, 0\}$ at the bottom
left corner, ${\bf q} = \{\pi, 0 \}$
at the bottom right corner, {\it etc}.
}
\end{figure}

The right panel in Fig.2 shows our result, where we superpose
the clean phase diagram and the case with 
A type disorder $\Delta_A=0.3$.
In the clean limit at $T=0$ as $\lambda/t$ is 
increased there is a transition
from a FM-M to the  A-2D phase at $\lambda/t \sim 1.52$,
and then a transition to a CE-CO-OO phase at $\lambda/t \gtrsim 1.55$.
On the FM-M side, $\lambda/t \le 1.52$, 
there is only a single thermal transition 
\cite{2d-mag}
at $T_C$ as one
cools the system.  At large $\lambda/t $, however, cooling 
first leads to a CO-OO phase, at $T_{CO}$, 
without magnetic order, followed by 
strong features in $S_{mag}$ 
at ${\bf q} = \{0, \pi \}$ and 
$\{0, \pi\}$, showing up at $T_{SR}$,
indicative of stripelike correlations. 
Finally,
at a lower $T$ the system makes a transition to
CE order.
If we set $t=0.3$eV, and use a factor of $3/2$ to convert
transition scales between 2D and 3D, our $T_C$ at
bicriticality would be $\sim 200$K.

In the presence of A type disorder with $\Delta_A =0.3$ we do
not find any spatial order on the CE side in either the 
charge, or orbital, or
magnetic sector, down to $ T \sim 0.005$.
The absence of order in the CE-CO-OO side 
can be traced back
to the `random field' $\epsilon_i$ coupling directly 
to the
charge order parameter $n_i$.
This breaks down charge correlations
to the atomic scale. The ferromagnet being a ${\bf q}=0$
state is more robust to A type disorder \cite{motome-prl}.

There are  short range stripelike magnetic
correlations that persist as peaks at ${\bf q} = \{0, \pi\}$
and 
$ \{\pi, 0\}$ in $S_{mag}({\bf q})$. The onset of this feature
is shown by the (red) dotted line in Fig.2.(b). This
appears even on the ferromagnetic side below
$T_C$. The $T_C$ itself is somewhat suppressed by
disorder and the ground state is an
{\it unsaturated} ferromagnet (u-FM).  Our analysis of 
the structure factor in the disordered system, however,
does not suggest any coexistence of
two distinct locally ordered phases at any $\lambda$. 
{\it A type disorder in the bicritical
regime does not induce phase coexistence.}
We have confirmed this directly from the
spatial snapshots as well, as we discuss later.

We have explored A type disorder with strength
$\Delta_A =0.1,~0.2,~0.3$ and $0.4$, over the range of $\lambda/t$
shown in Fig.2.(b). 
We now specialise to $\lambda/t=1.6$,
which is a CE-CO-OO phase near the clean phase boundary in Fig.2.(b)
and explore the  impact of A type and B type disorder in detail.
Fig.3.(a) shows the $T$  dependence of
the major peaks in the spin, charge and orbital structure 
factor in the clean limit at $\lambda/t=1.6$, for reference,
illustrating the  
distinct $T_{CO}$, $T_{SR}$ and $T_{CE}$ scales.

The naive expectation is that disorder would lead to
cluster coexistence \cite{dag-ps,nano-prl} 
of AF-CO phases, that arise for $x \ge  0.5$,
with the FM-M phase at $x \lesssim 0.4$, Fig.1.(b).
Fig.3.(c)  shows how the 
peaks in $S_{mag}({\bf q})$  evolve with
$\Delta_A$ at low temperature $(T=0.01)$. 
The peak at ${\bf q} = \{\pi/2, \pi/2\}$ vanishes
quickly, leading to a phase with stripelike correlations,
and the ${\bf q} = \{0, \pi\},~\{\pi, 0\}$ peaks also
vanish for  $\Delta_A >  0.6$ leaving a glass. 
The response to B type disorder is more interesting. We have
explored $V=1,~2$ and $4$ and $\eta=2,~4$ and $8\%$.
Since Cr is believed to be in a $t_{2g}^3e_g^0$ state
we focus here on $V=2$ which is sufficiently
repulsive to force $\langle n_i \rangle =0$ ($e_g^0$ state)
at the impurity sites. 
The response, as we 
vary the fraction of scatterers $(\eta)$, is 
similar to A type  at
weak $\Delta_B$. However, before the peak at 
${\bf q} = \{0, \pi\},~\{\pi, 0\}$ vanishes we see the emergence
of a 
peak at the ferromagnetic wavevector, ${\bf q}
= \{0, 0\}$. There is a window at intermediate $\eta$
where {\it B type disorder leads to coexistence of
FM and CO-OO-AF regions}.  In terms 
of transport, Fig.3.(b), intermediate A type disorder
strengthens the insulating
character in 
$\rho(T)$, while B type disorder of
comparable variance leads to an  insulator-metal
transition on cooling, and a (poor) 
metallic state at low temperature.

The top row in Fig.4 compares low temperature 
MC snapshots of
the magnetic correlations in the clean system at $\lambda=1.6$
(left), to that with $\Delta_A=0.5$ (center) and $\Delta_B=0.56$
(right).
The respective panels in the middle row
show the electron density
$\langle n_i \rangle$ corresponding to the panels above. The
panels at the bottom are the thermally averaged
$S_{mag}({\bf q})$ in the three cases.
In the clean limit
the magnetic correlations are CE, with  
a checkerboard density distribution, and
simultaneous 
magnetic peaks at ${\bf q} = \{0, \pi\},~\{\pi, 0\}
$ and $\{\pi/2, \pi/2\}$.
For A type disorder there are stripelike
magnetic correlations with small (atomic scale)
FM clusters but no signature of phase coexistence. The
density field is also inhomogeneous in the nanoscale, with
only short range
charge correlations, and 
$S_{mag}({\bf q})$ 
has weak peaks at ${\bf q} = \{0, \pi\}$ and $\{\pi, 0\}$
but no noticeable feature at ${\bf q}=\{0, 0\}$.  B type disorder, 
however,
leads to FM {\it regions} coexisting
with stripelike  AF correlations. The density field shows
a corresponding variation, being 
roughly homogeneous within the FM droplets (with
local density  $n \sim 0.6$), and a CO pattern away from the FM regions.
$S_{mag}({\bf q})$ now has peaks at
${\bf q} = \{0, \pi\},~ \{\pi, 0\}$ {\it and } $\{0, 0\}$,
as seen earlier in Fig.3.(d).

We explain the difference between the impact of A type and
B type disorder as follows. 
(1)~The introduction of A type
disorder does not lead to coexistence of large 
FM-M and AF-CO-OO clusters,
despite the presence of a PS
window in the clean problem, Fig.1.(b), 
because (a)~atomic scale
potential fluctuations disallow CO coherence beyond a few
lattice spacings, while (b)~homogeneous FM-M clusters are
destabilised by the disorder 
and become charge modulated. The result is a nanoscale 
correlated insulating 
glassy phase. (2)~Dilute strongly repulsive scatterers 
act very differently: (a)~they force an $e_g^0$ state at
the impurity sites and generate an `excess density' $0.5 \times
\eta$ which has to be distributed among the remaining
Mn sites,
(b)~the parent $x=0.5$ CO phase cannot accommodate this
excess charge homogeneously and the system prefers to phase
separate into $x \sim 0.5$ AF-CO and 
$x \sim 0.4$ FM clusters, (c)~unlike the A type
case, the FM clusters can survive and percolate since at
low $\eta$ there can be {\it large connected patches} without 
a B type site. We have verified this explicitly for several
impurity configurations. Making the
B site potential {\it strongly attractive} leads to a glassy
AF-CO state since carrier trapping 
reduces the effective electron count and forces the system
towards a combination of $x \ge 0.5$ phases in Fig1.(b).

In conclusion, we have 
reproduced all the key effects of A and B type disorder on 
phase competetion in the half doped manganites.
Our results suggest that B site impurities can
be chosen to engineer phase control 
and the percolative conduction paths
can be controlled through choice of 
dopant locations. 

We acknowledge use of the Beowulf cluster at HRI, 
and comments from E. Dagotto, S. Kumar and P. Sanyal.


{}


\begin{thebibliography}{99}
\bibitem{mang-book} 
See, {\it e.g}, {\it Colossal Magnetoresistive Oxides}, 
edited by Y. Tokura, Gordon and Breach, Amsterdam (2000).
\bibitem{ce-bicr1}
D. Akahoshi, {\it et al.}, Phys. Rev. Lett. {\bf 90}, 177203 (2003).
\bibitem{ce-bicr2}
R. Mathieu, {\it et al.}, Phys. Rev. Lett. {\bf 93}, 227202 (2004).
\bibitem{ph-comp-rev} 
Y. Tokura, Rep. Prog. Phys. {\bf 69}, 797 (2006).
\bibitem{cr-old1} 
A. Barnabe, {\it et al.}, Appl. Phys. Lett.  {\bf 71}, 3907 (1997).
\bibitem{cr-old2}
B. Raveau, {\it et al.}, J. Solid State Chem. {\bf 130}, 162 (1997).
\bibitem{cr-tok1}
T. Kimura, {\it et al.}, Phys. Rev. Lett. {\bf 83}, 3940 (1999).
\bibitem{cr-tok2}
T. Kimura, {\it et al.}, Phys. Rev. {\bf B62}, 15021 (2000).
\bibitem{cr-ps1} 
H. Oshima, {\it et al.}, Phys. Rev. {\bf B63}, 094420 (2001).
\bibitem{cr-ps2}
S. Mori, {\it et al.}, Phys. Rev. {\bf B67}, 012403 (2003).
\bibitem{class-approx} The validity of the classical approximations 
is studied in, {\it e.g}, E. Dagotto, {\it et al.}, Phys. Rev. 
{\bf B 58},
6414 (1998) and A. C. Green, Phys. Rev. {\bf B 63}, 205110 (2001).
\bibitem{brink-prl} 
J. van den Brink {\it et al.}, Phys. Rev. Lett.  {\bf 83}, 5118 (1999).
\bibitem{dag-ce-prl}
S. Yunoki, {\it et al.}, Phys. Rev. Lett. {\bf 84}, 3714-3717 (2000).
\bibitem{brey-prb} 
L. Brey, Phys. Rev. {\bf B 71}, 174426 (2005).
\bibitem{cepas} 
O. Cepas {\it et al.}, 
Phys. Rev. Lett. {\bf 94}, 247207 (2005).
\bibitem{dong-jt-prb}
S. Dong, {\it et al.}, Phys. Rev. {\bf B 73}, 104404 (2006).
\bibitem{ce-dag-03} 
H. Aliaga, {\it et al.},  Phys. Rev. {\bf B 68}, 104405 (2003). 
\bibitem{motome-prl}
Y. Motome, {\it et al.}, Phys. Rev. Lett. 91, 167204 (2003) 
\bibitem{ce-dis-dag}
G. Alvarez, {\it et al.},  Phys.  Rev. {\bf B 73}, 224426 (2006).
\bibitem{tca}
S. Kumar and P. Majumdar, Eur. Phys. J. {\bf B 50}, 571 (2006).
\bibitem{dag-ps}
Adriana Moreo, {\it et al.}, Phys. Rev. Lett. {\bf 84}, 5568 (2000).
\bibitem{nano-prl}
S. Kumar and P. Majumdar, Phys. Rev. Lett. {\bf 92}, 126602 (2004).
\bibitem{2d-mag} Our 2D magnetic ``$T_C$'' correspond to correlation 
length $\xi(T_C) \approx L$.  There is no genuine $T_C$ for
$L \rightarrow \infty$ in 2D.
The real 3D $T_C$ will be $\approx 3/2$ times
the 2D scale here.
\end{thebibliography}
\end{document}